\documentclass[11pt]{JHEP3}
\title{Finite Temperature effects on the Induced Chern-Simons term in
noncommutative geometry }
\author{Bhamidipati Chandrasekhar\\ Institute of Physics, Sachivalaya Marg,\\
Bhubaneswar 751 005, India\\ E-mail \email{chandra@iopb.res.in}}
\author{Prasanta K. Panigrahi\\ School of Physics, University of
Hyderabad,\\ Hyderabad 500046, India, and \\ 
Physical Research Laboratory,\\ Navrangpura, Ahmedabad 380009, 
India \\E-mail 
\email{prasanta@prl.ernet.in}}
\abstract{The one-loop contribution to vacuum polarization  is
calculated for the adjoint fermions in three dimensional noncommutative
spaces, both at zero and finite temperature.  At zero temperature, we
confirm a previously found result for the parity odd part and subsequently 
analyze the even parity part, which exhibits UV/IR mixing.
We discuss in detail, two
regimes of the high temperature behavior of the parity odd part. When
the thermal wavelength is much smaller, as compared to the noncommutativity
scale, we find an interesting Fermi-Bose transmutation in the nonplanar part.}
\keywords{Chern-Simons Theories, Non-Commutative Geometry, Thermal Field
Theory} 
\preprint{IP/BBSR/2003-2\\
hep-th/0301030\\} 
\begin{document}
\begin{section}
{\bf {Introduction}}

The study of field theoretical models on noncommutative spaces has
attracted considerable attention in recent times (for a review see,
\cite{Nekrasov}), because of their natural appearance in certain limits
of the string and M theories, in a constant background B field. The
inverse of $|B|$ plays the role of the dimensionful noncommutativity
paramater $\theta$ \cite{Witten,Connes}. Noncommutative theories show very
interesting perturbative behavior \cite {Minnwalla,Raamsdonk,Toumbas,
Sondhi}, due to the presence of an extra Moyal phase at the vertex. In a
simple model, this phase has been shown to arise due to the interaction of
pairs of oppositely charged particles, restricted to the lowest Landau
level (LLL), by a strong magnetic field \cite{Bigatti,Jabbari}. Since the
noncommutative field theory describes particles, which are extended like
dipoles, it has been proposed that, the long distance behavior of the
quantum Hall fluid \cite{Girvin} may be better modelled by a
Chern-Simons (CS) theory \cite{Jackiw,Dunne} in a noncommutative
geometry \cite{Susskind,Polychronakos}.

CS theories on noncommutative spaces are being extensively
studied \cite{Jabb,Chen,Khare,Nishimura,Martin,Frohlich,Wu,Gorsky} 
due to their potential applications in
condensed matter systems. A number of formal aspects of these theories, in
parallel to their commutative counter parts, are also being
investigated \cite{stern,poly}. It has been shown that the 
coefficient of the CS term
remains quantized for the non-abelian and, surprisingly, also for the
abelian theory, on a noncommutative space \cite{Bak,Nair}.  Using the
arguments of BRST invariance and linear vector supersymmetry, the
Coleman-Hill theorem \cite{coleman} has been extended to the
noncommutative background, showing that the tree level CS term does not
receive any more correction, beyond one loop, in pure CS gauge theories
\cite{Das,Bichl}.

The vacuum polarization tensor, at the one-loop level, is not affected by the
noncommutative geometry, for the fermions in the fundamental
representation. Charged fermions, in the adjoint representation induce a
~$\theta$~ dependent CS term, which shows a discontinuous behavior in the
noncommutativity parameter~ $\theta$ \cite{Chu,grandi}.

In the ordinary CS theory, the effect of finite temperature, particularly
on the parity odd sector \cite{pkp}, leads to a number of interesting results
\cite{large}. The parity odd part of the photon self-energy loses its
analyticity at finite temperature \cite{kao,aitch}. It has been pointed out
that, invariance of the effective action under large gauge
transformations, at non-zero temperatures, necessitates the incorporation
of nonperturbative contributions \cite {dunne}. Apart from these formal
aspects, physical applications of the CS theories in noncommutative
geometry, make it imperative to study these theories at finite temperature.

In this note, we study the effect of temperature on the induced CS term
for the adjoint fermions.  The 
parity odd part of the polarization tensor is analyzed carefully. Various
limits, involving the noncommutativity paramater and the high and 
low temperature behaviors 
of the coefficient of
the induced CS term are studied in detail and a number of interesting
features pointed out. The analysis of the even parity sector reveals UV/IR
mixing, a hall-mark of the noncommutative theories.

The paper is organized as follows. In section II, we outline the general
features of field theories on noncommutative spaces and then proceed to
calculate the one-loop vacuum polarization tensor. Both the odd and
even parity sectors are studied and the results contrasted with those 
of the 
ordinary CS theory.  In section III, the finite temperature effects are
analyzed and various limiting behaviors are studied in detail. We
conclude in section IV, after pointing out directions for future study.
\end{section}

\begin{section}
{\bf {Field theories on noncommutative spaces}}
There are two equivalent descriptions of the noncommutative spaces (for a
lucid review, see \cite{Micu}), which is taken as $2+1$ for our purposes.
When the
coordinates of the space are considered as operators, noncommutative
$R^{3}$ is defined by the fundamental relation,
\begin{equation}
[X_{i} , X_{j}] = i \theta_{ij} ~~~,
\end{equation}
where $\theta$ \footnote{Henceforth, we only consider the case where
$\theta _{0i} = 0.$} is a constant antisymmetric matrix, carrying a
dimension of the square of length. Hence, functions on $R^{3}$ would be
functions of operators. However, a more convenient way is to work with
functions of real variables, but with the ordinary product replaced by the
Moyal star product. In this description, (2.1) takes the form,
\begin{equation}
[x_{i} , x_{j}]_{MB} = i \theta_{ij} ~~~, 
\end{equation}
where a Moyal bracket (MB) between two functions f(x) and g(x) is defined as,
\begin{equation}
[f(x) , g(x)]_{MB} = (f * g)(x) - (g * f)(x) ~~~. 
\end{equation}
The associative star product is defined as,
\begin{equation} 
(f*g)(x) = [exp(\frac{i}{2} \theta_{\mu \nu}
\partial_{\alpha_{\mu}} \partial_{\beta_{\nu}})f(x + \alpha) g(x+
\beta)]_{\alpha=\beta=0} ~~~.
\end{equation}
Notice that the star product, defined at a point, brings in nonlocality
due to the presence of infinite number of derivatives.  A great
simplification, while carrying out perturbative expansions in a
noncommutative field theory, occurs due to the following result:
\begin{equation}
\int (f*g)(x) d^{3}x = \int (g*f)(x) d^{3}x = \int f(x)g(x)d^{3}x ~~~.
\end{equation}
When the interaction is switched off, the results of the ordinary theory
go over to those of the noncommutative theory. The Green's functions are
not modified and the perturbative calculations proceed along similar lines
as those of the oridinary cases. However, the vertices carry additional
phases originating from the Moyal bracket. Hence, at the quantum level,
two types of diagrams need to be analyzed: planar and nonplanar. The
planar diagrams are those, where the phase factor depends on the external
momenta and the nonplanar diagrams are the ones, where the phase factor
contains internal loop momentum. This is responsible for a number of
interesting features of the noncommutative theories.\\

The QED action (for the construction of QED
action and the
Feynman rules, see \cite{Hayakawa}) for the 
two component fermionic field is given by,\\
\begin{equation} 
S[A,m] = \int d^{3}x[ -\frac {1}{4} F_{\mu \nu}* F^{\mu
\nu} + (\overline \psi *(i D\!\!\!\!/ - m)\psi)(x) ] ~~~, 
\end{equation}
where m is the bare mass of the fermion. The field strength tensor,\\
\begin{equation}
F_{\mu\nu} = \partial_{\mu} A_{\nu} -  \partial_{\nu} A_{\mu}  - i g [A_{\mu}
 , A_{\nu}]_{MB} ~~~,
\end{equation}
reveals that even in a U(1) theory, the gauge field $A_{\mu}$ is
self-interacting, with the coupling constant g. It can be checked that the
above action is invariant under the star gauge transformations
\footnote{Although, the divergence of the current does not vanish ( a
consequence of the star product), it can be shown that a conserved charge
does exist.} $\delta \psi = ig\alpha*\psi$ for the
Dirac spinor (fundamental representation) and $\delta \psi = ig[\alpha ,
\psi]_{MB}$ for the Majorana spinor (adjoint representation). Here, the
covariant derivative, for the Dirac fermion is $D_{\mu} \psi =
\partial_{\mu}\psi - ig A_{\mu}*\psi $ and for the Majorana spinors, it is
given by, $D_{\mu} \psi = \partial_{\mu}\psi - ig [A_{\mu},\psi]_{MB} $~;
in the limit ~$\theta \rightarrow 0$~, the Majorana coupling to
the gauge field vanishes. \\

For convenience of comparison with the existing results, we work in the 
Euclidean space in 2+1 dimentions. The
relevant Dirac gamma matrices, satisfying, $\{\gamma^{\mu},\gamma^{\nu}\}=
-2\eta^{\mu\nu} $~, are given by,
$\gamma^{1}= i\sigma^{2}$ ~,~ $\gamma^{2}= i\sigma^{3}$~,~
$\gamma^{3}= i\sigma^{1}$ ~,~
 and some
useful identities needed for later use are,\\
$~~~~~~~~~~~~~~\gamma^{\mu}\gamma^{\nu} = -\eta^{\mu\nu} - 
\epsilon^{\mu\nu\lambda}\gamma^{\lambda}$ ~,~~~
$Tr(\gamma^{\mu}\gamma^{\nu}\gamma^{\lambda}) =
2\epsilon^{\mu\nu\lambda}$ ~~~,\\
$Tr(\gamma^{\mu}\gamma^{\nu}\gamma^{\lambda}\gamma^{\rho}) =
2(\eta^{\mu\nu}\eta^{\rho\lambda} + \eta^{\mu\rho}\eta^{\nu\lambda} -
\eta^{\mu\lambda}\eta^{\nu\rho}) ~~~. $ \\

As has been mentioned earlier, the quadratic part of the action is the
same as that of the ordinary theories and hence, the perturbative 
calculations can be carried out analogously. From the
expansion of the action in the momentum space, it follows that, the vertex
contains an additional phase factor $e^{i\tilde p k}$, as compared to the
commutative case, here $ \tilde p k= p_{i} \theta ^{ij} k_{j} $.

\begin{subsection}
{Vacuum polarization in 2 + 1 dimensions}

In this section, we present the calculations for the vacuum polarization
tensor, separating out the planar and nonplanar contributions. The odd
and even parity sectors are discussed in appropriate
subsections.  The photon self-energy gets modified due to the presence of
the Moyal phases at the vertices and is given by,\\
\begin{equation} 
i\Pi^{\mu\nu} (p_{3},\overline p)= -g^2 \int \frac
{d^{3}k}{(2\pi)^3} \frac {{ \mathit {M} }^{\mu\nu} }{((k-p)^2 + m^2 
)(k^2 + m^2 )} . (V) ~~~, 
\end{equation}
with,\\
\begin{equation}
{\mathit {M} }^{\mu\nu} = 2(2k^{\mu} k^{\nu} - (p^{\mu}k^{\nu} +
p^{\nu}k^{\mu}) + \eta^{\mu\nu} ( p.k - k^2) - m^2 \eta^{\mu\nu} + m
p^{\alpha}\epsilon^{\mu\nu\alpha}) ~~~.
\end{equation}
The only change in going to the noncommutative theory is the appearance
of the Moyal phase V \cite{Filk}; this turns out to be one, for the case
of the Dirac fermions and $ 4~sin^2(\frac {\tilde p k}{2})$
for the Majorana case. The dependence of
V on the internal loop momentum implies that the polarization tensor can
receive contributions from the non-planar diagrams as well.  \\

The case of the Dirac fermions needs no further elaboration, as it
reproduces the results of the ordinary theory; henceforth, we shall
concentrate on the Majorana case. Writing, 
$sin^2(\frac {\tilde p k}{2}) = \frac{1}{2} (1- cos( \tilde p k) )$, 
we can separate out the
contributions of the planar and non-planar parts to the polarization
tensor. Making use of the Feynman parameterization, one obtains,
\begin{equation}
i\Pi^{\mu\nu}(p) = - 2 g^2 \int _{0}^{1} ~dx \int
\frac{d^3l}{(2\pi)^3}[\frac {1}{(l^2 + M^2)^2} - \frac{1}{2} \frac { (
e^{\tilde p l} + e^{-\tilde p l} )}{(l^2 + M^2)^2} ]{\mathit {M}
}^{\mu\nu} ~~~,
\end{equation}
with, $M^2 \equiv -x(x-1)p^2 + m^2$~~ and\\
\begin{eqnarray*}
{\mathit {M} }^{\mu\nu} &=& 2[2 l^{\mu}l^{\nu} + (2x-1)( l^{\mu} p^{\nu} +
l^{\nu} p^{\mu}) - \eta^{\mu\nu}(l^2 + (2x-1)l.p) \\ 
&+& 2x(x-1) p^{\mu}p^{\nu} - \eta^{\mu\nu}x(x-1)p^2 - m^2\eta^{\mu\nu} +
mp^{\alpha} \epsilon^{\mu\nu\alpha}] ~~~. \\
\end{eqnarray*}
The planar part can be readily evaluated, making use of the Pauli-Villars
regularization; for the purpose of comparison, it is better to evaluate
both the planar and nonplanar parts in the same regularization scheme, as will
be done below. Exponentiating the denominator using the familiar Schwinger
parameterization: $\frac {1}{k^2} = \int_{0}^{\infty} d\alpha
~exp(-\alpha k^2)$~, we get,\\
\begin{eqnarray*}
i\Pi^{\mu \nu}(p) &=& -2g^2 ~\int dx \int d\alpha \int \frac {d^3~l }
{ (2\pi)^3 } ~2[{\mathit {M} }^{\mu\nu}_{P} - {\mathit {M} }^{\mu\nu}_{NP}.~
e^{ \frac {-\tilde p^2 }{4\alpha} }]\\
&\times& ~\alpha~.exp~\alpha(-l^2-M^2) ~~~,\\
\end{eqnarray*}
wherein,\\
\begin{equation}
{\mathit {M} }^{\mu\nu}_{P} = (2 l^{\mu}l^{\nu}  - \eta^{\mu\nu}l^2)
+  x(x-1)(2 p^{\mu}p^{\nu} -\eta ^{\mu\nu}p^2)
- m^2\eta^{\mu\nu} + mp^{\alpha} \epsilon^{\mu\nu\alpha} ~~~.\\
\end{equation}
Below, we give the components of the non-planar tensor part, which has
additional terms:\\
\begin{equation}
{\mathit {M} }^{\mu\nu}_{NP} = {\mathit {M} }^{\mu\nu}_{P} + 
{\mathit {M} }^{\mu\nu}_{n} ; 
\end{equation}
the components of the additional tensor part turn out to be:\\
\begin{equation}
{\mathit {M} }^{00}_{n} = - \frac {\tilde p^2}{4\alpha^2}
,~~~{\mathit {M} }^{11}_{n} =  \frac {\tilde p^2}{4\alpha^2} - 
\frac {\tilde p_2^2}{2\alpha^2},~~~ 
{\mathit {M} }^{22}_{n} =   \frac {\tilde p^2}{4\alpha^2} - 
\frac {\tilde p_1^2}{2\alpha^2} \\
\end{equation}
\begin{equation}
{\mathit {M} }^{01}_{n} = {\mathit {M} }^{02}_{n}= 0  
,~~~and~~~ {\mathit {M} }^{12}_{n}= \frac {\tilde p_1\tilde p_2}{2\alpha^2}. \\
\end{equation}


The extra terms in the nonplanar part appear due to the 
momentum redefinition.
\end{subsection}

\begin{subsection}
{The odd parity sector}

The parity odd part of the vacuum polarization tensor gives rise to the
induced CS term.  
Carrying out the momentum integration, one finds,\\
\begin{equation}
i\Pi^{\mu \nu}_{odd}(p) = -\frac {mg^2}{2 \pi^{ \frac{3}{2} } } p^{\alpha}
\epsilon^{\mu \nu \alpha}\int dx \int \frac { d\alpha}{\sqrt {\alpha}} 
~[ e^{-\alpha M^2} - e^{ -\alpha M^2 - \frac {\tilde p^2}{4\alpha
}}] ~~~.
\end{equation}
The integrals are finite and the $\alpha$ integration can be done 
straightforwardly:\\
\begin{equation}
i\Pi^{\mu\nu}_{odd}(p) =  -\frac {mg^2}{2 \pi^{\frac {3}{2} } } p^{\alpha}
\epsilon^{\mu \nu \alpha} \int dx [ \frac {\sqrt {\pi}}{M} - \sqrt {\frac 
{2\tilde p}{M} }~{\mathit {K}}_{\frac {1}{2}}(\tilde p~M) ] ~~~,
\end{equation}
where, ${\mathit {K}_{\frac {1}{2}}(\tilde p~M)}$ is the modified Bessel
function of the second kind.  
Using,\\
\begin{equation}
{\mathit {K}}_{\frac {1}{2}}(z) = \sqrt {\frac {\pi}{2z} }~e^{-z} ~~~,
\end{equation} 
we end up with,\\
\begin{equation}
i\Pi^{\mu\nu}_{odd}(p) =  -\frac {mg^2}{2 \pi } p^{\alpha}
\epsilon^{\mu \nu \alpha}\int dx [ \frac {1}{M} -  \frac {1}{M}~
e^{- \tilde p~M }] ~~~,
\end{equation}
as has been noticed earlier \cite{Chu}. In the lowest order in p, the planar
part gives rise to the the first term in the following CS term, 
in the effective action:\\
\begin{equation}
W[A] = \frac{m}{|m|} \frac{g^2}{4\pi} \int d^3x
[~\epsilon^{\mu\nu\lambda}(A_{\mu}\partial_{\nu}A_{\lambda}
+ \frac{2i}{3}g A_{\mu}*A_{\nu}*A_{\lambda})~] ~~~.\\ 
\end{equation}
Notice that the cubic term in the above expression can be derived by 
calculating the 3-point function, where the net phase factor at the vertices
does not vanish and depends only on the external momentum,
accounting for the star product.\\

There exists 
the following two physically interesting limits:

{\bf {In the limit~ $\theta \rightarrow 0$~}}, the planar and
non-planar contributions cancel each other. This is expected, since the
Majorana fermions do not couple to the photons in the ordinary theory.
Hence, this turns out to be the ordinary theory limit.\\

{\bf {The limit ~$m \rightarrow \infty$~}} keeping ~ $\theta$~
constant. This situation arises, as will be seen later, when the regulator 
field is introduced. It
is this limit in the ordinary theory that generates the CS term. Here, we
see that, the non-planar contribution vanishes leaving the first term as
the induced CS term. The large m limit also corresponds to the extreme
noncommutativity limit, in which, all the non-planar contributions vanish
and one is only left with the planar diagrams.

The observation that the above two limits, ~ $\theta \rightarrow 0$~ and
~$M \rightarrow \infty$~ do not commute
has been termed as the discontinuity in ~ $\theta$ ~\cite{Chu}. 
\end{subsection}

\begin{subsection}
{Parity even part}

Unlike the parity odd case, it can be easily seen that the momentum
integration here, gives rise to a divergent contribution, needing
regularization. The main purpose being the identification of UV/IR mixing,
we focus on the trace of the self-energy. Using the identities given
earlier and carrying out the $l$ and $\alpha$ integrations, we obtain
(see appendix),\\
\begin{eqnarray*}
i\Pi^{\mu \mu}_{even}(p)& =& g^2 \int d~x [~ \{~ \frac {3 \Lambda}{2\pi^2} 
~e^{- \frac {M}{\Lambda} } + \frac {1}{2M \pi^2}~(- x(1-x)p^2 + 3m^2)~ \}
\\
&-& \{~ \frac {3 \Lambda_{eff} }{2\pi^2}
~e^{- \frac {M}{\Lambda_{eff} } } + \frac {1}{2M \pi^2}~(- x(1-x)p^2 
+ 3m^2)
~e^{- \tilde p M }~ \}
\end{eqnarray*}
\vskip -0.3in
\begin{eqnarray}
~+ \{~ \frac {M}{2\pi^2}~ \Lambda_{eff}^2~(1 + \frac { \Lambda_{eff} }{M} )
\tilde p^2~\}] ~~~,
\end{eqnarray}
where a cut-off parameter $\Lambda$ has been introduced to take care of
the small $\alpha$ divergence. Here, $\Lambda_{eff}^{-2} = \tilde p^2 +
\frac {1}{\Lambda^2}$~ is the effective cut-off. In the above expression,
the first term represents the planar contribution, where, the divergent
and the gauge invariant pieces have been isolated. The second and third
terms are the non-planar contributions.

Notice that even in the absence of the
cut-off~$\Lambda$, the non- planar contribution receives a natural cut-off
due to the non-commutativity paramater, which makes the integrals UV
finite. Like the odd parity case, various limits yield a number of 
interesting behaviors.

{\bf {$\tilde p \ll 1/\Lambda$}}, corresponds to taking the
$\theta \rightarrow 0$ limit, and $\Lambda_{eff} \equiv \Lambda $. We see
that dropping the $\theta$ dependent terms, the planar and non-planar
contributions cancel each other, reproducing the result of the ordinary
theory.

{\bf {$\tilde p \gg 1/\Lambda$}}, corresponds to the limit
$\Lambda \rightarrow \infty$,~ with $\Lambda_{eff}^{-1} \equiv \tilde p$ and
we end up with a piece $\frac {2}{\tilde p}~ \times~ finite$, where we
have neglected a divergent part in the planar piece, which is exactly
cancelled by the regulator field. We see that the limit~ $\tilde p
\rightarrow 0$~ is IR singular. Although, ~$\theta$~ regulates the
integrals by providing a UV cut-off, it produces a new IR singularity.
This is the UV/IR mixing, a characteristic feature of the noncommutative
theories \cite{Raamsdonk}. The physical implications of the above
IR singularity need further study. This is of interest for theories 
dealing with LLL physics
and other cases involving phase transitions in 2+1 dimentions. It would be
interesting to look at the above scenerio at finite temperature, as it is
known from the ordinary theories that, temperature effects often bring in
IR divergences as well \cite{Ashok}. However, we shall confine ourself 
here to the
study of finite temperature effects on the parity odd part of the
polarization tensor, leaving the above analysis for a future work.

\end{subsection}
\end{section}


\begin{section}
{Finite temperature effects on the induced Chern-Simons term }

In this section, we explore the effect of finite temperature on the
one-loop parity odd part of the polarization tensor. We take recourse to
the widely used method in thermal field theory, the Imaginary time 
formalism for studying the equilibrium
systems, where the time variable is given up in favour of 
temperature \cite{Ashok,Kaputsa}.

In ordinary theories, it is known that some amplitudes at finite 
temperature are
nonanalytic functions of the argument $p$, i.e., the quantities $\Pi(p_0 =
0, \overline p \rightarrow 0)$ ~and ~$\Pi(p_0 \rightarrow 0, \overline p =
0)$ do not yield identical results. Here, we would be interested
in calculating the odd parity part only in the static limit $(p_0 = 0,
\overline p \rightarrow 0)$ ~, as in the other limit, the effect of
noncommutativity is absent, to start with.

Just like in the ordinary theory, we shall take over the expressions at
zero temperature, and replace the continuous energy variable $k_{3}$ by
$\frac {2\pi n}{\beta}$ for bosons and 
$\frac {2\pi }{\beta}(n + \frac{1}{2})$
for fermions, and
the integration over this variable is replaced by a discrete sum: $\int
(dk_3/2\pi) ~~\rightarrow ~ \frac {1}{\beta} \sum_{n}$.  
Hence, the vacuum
polarization tensor in the static limit for the fermions in the adjoint
representation at finite temperature is:
\begin{equation}
i\Pi^{\mu\nu}_{(\beta,\mu)} (0,\overline p) = -\frac {4g^2}{\beta} \sum_{n}
\int \frac{d^2k}{(2\pi)^2} \frac { {\mathit {M} }^{\mu\nu}. sin^2(\frac
{\tilde p k}{2}) }{( (k-p)^2 + m^2 + \omega_{n}^2)(k^2 + m^2 +
\omega_{n}^2 )} ~~~,
\end{equation}
At finite temperature, the most general expression for the polarization
tensor can be written as: 
\begin{equation}
\Pi^{\mu\nu} = \Pi^{\mu\nu}_{even}(p) - \epsilon^{\mu\nu\alpha} p^{\alpha}
~\Pi^{odd}(p) ~~~. 
\end{equation}
We would be interested in calculating the one-loop contribution to the
parity odd part, $~\Pi_{odd}(p)$, which gives rise to the induced CS term.
It is convenient to do the
summation first, in order to get an expression in a closed form. Hence,
after Feynman parameterization, we have the planar and non-planar pieces
isolated as:
\begin{equation}
\Pi_{odd}(p) = -\frac {4mg^2}{\beta}
\sum _
{n=- \infty}^{\infty
}~\int_{0}^{1} dx \int \frac {d^2l}{(2\pi)^2} [\frac {1 - \frac{1}{2}\{e^{il
\tilde p} + e^{-il\tilde p} \} }{(M_1^2 + ((2n+1)\frac {\pi}{\beta} +i\mu )^2
}~] ~~~, 
\end{equation}
where $M_1^2 = \overline l^2 -x(1-x)p^2+ m^2$~ and ~$\mu$~ is the chemical
potential. 
The summation can be done
by standard methods. We note that, both the exponentials 
in (3.3) contribute the
same.  Going to polar coordinates and setting $p_1 = 0$~for
convenience, we are left with,
\begin{eqnarray*}
\Pi_{odd}(p) &=& \frac {-ig^2}{2\pi^2}~ \frac 
{\partial}{\partial m} \int_{0}^{1} dx \int_{0}^{2\pi} \int_{0}^{\infty} 
\frac {rdr
d\phi}{(r^2 + M^2)^{\frac {1}{2}}}~[~2~ -~ \frac {2}{e^{\beta \sqrt 
{r^2 + M^2} +\mu} +1 }\\
&-&~  \frac {2}{e^{\beta \sqrt 
{r^2 + M^2} -\mu} +1 }~][~1~ -~ e^{ir\tilde p_2 cos\phi}~] ~~~,\\ 
\end{eqnarray*}
where as before, we have defined, $M^2 = -x(1-x)p_2^2 + m^2$.  The
calculation of the planar part is straightforward,
yielding the following temperature dependence for the
coefficient of the CS term for the case of ~$\mu = 0$~,
\begin{equation}
\Pi_{odd}^{P}(0) = \frac {2ig^2}{\pi}
\frac{m}{|m|} ~tanh(\frac {\beta m}{2}) ~~~. 
\end{equation}
This is the same result as one gets in the ordinary theory with
fundamental fermions. It is worth
noting that, the above result can be generalized to selective field
configurations, where the full effective action depends on temperature in
a manner, that preserves the large gauge invariance \cite{Rossini}.

The result of integration for the nonplanar piece using the identities
given in the appendix is:
\begin{eqnarray*}
\Pi_{odd}^{NP}(p)
= -\frac {2ig^2}{\pi}~ \int_{0}^{1} dx~ 
~\frac {m}{|M|}~
[~e^{-M\tilde p_2}~ -~ 2
~\{2~ e^{-M\sqrt {\beta^2 + \tilde p_{2}^{2} }}  
\end{eqnarray*}
\vskip -0.3in
\begin{eqnarray}
~~~~~~~~~~~~~~~~~~~~~~~~~~~~~~~~~~~~~~~~~~~~~~~~~~~~~~
- \sum_{n = 1}^{\infty}~ 
e^{-M\sqrt {n^2\beta^2 + \tilde p_{2}^{2} }}~
\} ~] ~~~,
\end{eqnarray}
wherein, the summation over n in the above expression appears upon using the
result, ~$\frac {1}{e^{x} + 1} = 2~e^{-x} - \sum _{n=1}^{\infty} e^{-nx}$.
At this stage, we can check that the limit ~$\theta \rightarrow 0$~ is 
smooth, since,
\begin{equation}
\Pi_{odd}^{NP}(0) =
-\frac {2ig^2}{\pi}\frac {m}{|m|}
tanh(\frac {\beta m}{2}) ~~~,
\end{equation}
which is exactly the negative of (3.4). As is
obvious, the planar and the non-planar parts cancel even at finite
temperature. Also, just
like at zero temperature, the nonplanar contribution vanishes in the large
m limit. Hence, the discontinuous behaviour of $\theta$ persists even at
finite temperature.

Since, at finite temperature we have an extra dimensionful parameter 
~$\beta$~, two
different regimes of high temperature can be studied: 
~$p\theta/\beta\ll 
1$~, and ~$p\theta/\beta \gg 1$~ similar to the limits considered
by
\cite{Lopez,Arcioni,fishler}. We also find that the limit 
~$p\theta/\beta~ \ll 1$~ 
produces the results of the conventional field theory. In our case, this
limit corresponds to taking ~$\theta \rightarrow 0$,~ keeping m fixed,
which, as we have checked, indeed reproduces the ordinary theory results.

The other regime of high temperature, ~$p\theta/ \beta \gg 1$~, 
where the thermal wavelength is much smaller than the
noncommutativity scale shows some interesting features. 
To find out the behavior of the nonplanar part
in this regime, we need to perform the remaining summation in (3.5), 
which can be done using the Poisson summation formula:
\begin{equation}
\sum _{n=1}^{\infty}~e^{-M\sqrt {n^2\beta^2 + \tilde p_{2}^{2} }} =
-\frac{1}{2}~e^{-M\tilde p_2} + \sqrt {2\pi}~[~\frac {1}{2}~F(0) + 
\sum _{n=1}^{\infty} F(2\pi n)] ~~~,
\end{equation}
with the function $F(2\pi n)$ given by,\\
\begin{eqnarray*}
F(2\pi n) &=& \sqrt {\frac{2}{\pi}} \int _{0}^{\infty} dt~ 
e^{-M\sqrt {t^2\beta^2
+ \tilde p_{2}^{2}}}~cos(2\pi nt)\\
&=& - \sqrt {\frac{2}{\pi}}~\frac {1}{\beta}~\frac {\partial}{\partial M}~
{\mathit {K}}_{0}
(\frac {\tilde p_2}{\beta}~\sqrt {4\pi^2 n^2  + M^2\beta^2}) ~~~.
\end{eqnarray*}
Here, we have made use of the result \cite{grad},
\begin{eqnarray*}
\int _{0}^{\infty} \frac {dx}{\sqrt {b^2 + x^2} }~e^{-a~\sqrt {x^2 + b^2} }
~cos(tx) = {\mathit {K}}_{0}(b\sqrt {t^2 + a^2}) ~~~,
\end{eqnarray*}
${\mathit {K}}_{0}$ being the modified Bessel's function of the second
kind. Now, we are left with the summation of $F(2\pi n)$ which seems
problematic due to the presence of a squareroot in the expression.
However, using the following integral representation for ${\mathit
{K}}_{0}$,
\begin{equation}
 {\mathit {K}}_{0}(z) = \frac {1}{2} \int _{0}^{\infty} \frac{dt}{t}
 ~e^{-\frac {1}{2}(t + \frac {z^2}{t}) } ~~~,
\end{equation}
it is possible to get rid of the squareroot, converting the summation in
to a Gaussian form. Note that the change of order of summation and integration
is not justified at ~$z=0$~, 
where the above integral blows up. Thus, we
\begin{equation}
\sum _{n=1}^{\infty} F(2\pi n) = - \sqrt {\frac{2}{\pi}}~\frac {1}{2\beta}
~\frac {\partial}{\partial M}~ \int_{0}^{\infty} \frac {dt}{t}~
e^{ -\frac {1}{2}(~t ~+ \frac {M^2 \tilde p_{2}^{2}~}{t})  } ~
\sum _{n=1}^{\infty}e^{-\frac {2\tilde p_{2}^{2} \pi^2}{\beta^2 t}~n^2}~~.\\
\end{equation}
In the limit ~$p\theta/\beta >> 1$~, the temperature dependent terms
in the above equation are exponentially damped. The above can be
evaluated term by term. Considering the leading order contribution,
we get:
\begin{equation}
\sum _{n=1}^{\infty} F(2\pi n)~ \approx ~\frac{T~ \sqrt {M^2\tilde p_2}}
{(~M^2 + (2\pi T)^2~)^{3/4}}~ e^{-\tilde p_2 \sqrt {~M^2 + (2\pi T)2~}}~~~.\\
\end{equation}
Using (3.10), the summation in (3.7) can be written as,\\
\begin{eqnarray*}
\sum _{n=1}^{\infty}~e^{~-M\sqrt {n^2\beta^2 + \tilde p_{2}^{2}}} ~\approx~
~-~\frac {1}{2}~ e^{M\tilde p_2} ~+~ \frac {\tilde p_2}{\beta}~ 
K_1(M\tilde p_2)
\end{eqnarray*}
\begin{eqnarray}
~~~~~~~~~~~~~~~~~~~~~~~~~~~~~~~~~~~~~~~~~~~~~~~
~+~ \frac{T~ \sqrt {M^2\tilde p_2}}
{(~M^2 + (2\pi T)^2~)^{3/4}}~ e^{-\tilde p_2 \sqrt {~M^2 + (2\pi T)2~}}~.
\end{eqnarray}
Using (3.11) in (3.5), the nonplanar contribution
to the parity odd part for the regime ~$p\theta/\beta \gg 1$~
comes out to be:\\
\vskip -0.2in
\begin{eqnarray*}
\Pi_{odd}(p)~ \approx~
~-~ \frac{4ig^2}{\pi}~\int_{0}^{1} dx~\frac{m}{|M|}
~[~~\tilde p_2~ T~ K_1(\tilde p_2 M)~-~2~ e^{-M \tilde p_2} ~~~~
\end{eqnarray*}
\begin{eqnarray}
~~~~~~~~~~~~~~~~~~~~~~~~~~~~~~~~~~~~
~+ ~\frac {T~\sqrt {~2\pi M^2 \tilde p_2}~}
{(~M^2 ~+~ (2\pi T)^2~)^{3/4}}~ e^{-\tilde p_2 \sqrt {~M^2 + (2\pi T)2~}}
~~]~~~~.
\end{eqnarray}
We note that the leading order behavior of the above expression is linear 
in T.
The nonplanar piece shows a peculiar high temperature behavior, very
different from the planar piece, which goes as ~$\frac {1}{T}$~. The linear
T dependence of the non-planar piece is similar to the behavior of
bosonic loops in ordinary theories at finite temperature, in the static
limit \cite{ashokdas}.  

It is interesting that for thermal wavelengths larger than the
noncommutativity scale, the nonplanar contribution to the odd
parity part of the polarization tensor shows the temperature dependence
characteristic of fermions and for wavelengths smaller than the ~$\theta$~
scale, a temperature dependence characteristic of bosons.
Evidence from other works \cite{fishler}, regarding these two regimes of
high temperature for the nonplanar part suggests that, the system undergoes
some kind of a phase change once the temperature is raised beyond
~$\sqrt {\theta}$~. Clearly, the regime 
~$p\theta/\beta~ \gg 1$~ needs a
better understanding.

Also, the above result for the non-planar piece is not large gauge
invariant. In the case of ordinary theory, 
the full effective
action for selective, time independent backgrounds has been shown to be
invariant under large gauge transformations at finite temperature 
\cite{salcedo}.
Hence, it is worthwhile to 
perform a similar
non-perturbative computation in this context as well. 
\end{section}

\begin{section}
{\bf {Conclusions}}
To conclude, the CS term induced from the fermions in the fundamental
representation does not receive any corrections due to the
noncommutativity of space, both at zero and finite temperature, at one
loop. However,
the CS term induced from the adjoint fermions at zero temperature,  
shows a
discontinuous behavior as a function of the noncommutativity parameter,
which
persists even at finite temperature.  The parity even part shows UV/IR 
mixing. There are two regimes of
high temperature one can distinguish. The regime 
~$p\theta/\beta \ll 1$~,
produces the results of ordinary theory, whereas in the regime 
~$p\theta/\beta \gg1$~,
the nonplanar contribution to the odd parity part of the
polarization tensor shows a very different temperature dependence;
the behavior is similar to that of bosonic loops in
ordinary theories, whereas the planar piece still goes as ~$1/T$~,
a feature characteristic of the behavior of fermionic loops.
Evidently, the contrasting high temperature behaviors of the planar and
non-planar pieces in the above regime need further study. 

The computation
of the effective action invariant under large gauge transformations, 
should throw more light on this aspect. Also, it
would be worthwhile to look at the even parity part in the above two regimes
of high temeprature as would be the study of the polarization tensor at finite
density, even for the non-abelian case.
These works are in progress and we hope to get back to these issues in future.\\
\end{section}

\acknowledgments

B.C. would like to thank the warm hospitality of School of Physics, 
University of Hyderabad and Institute of Mathematical Sciences, Chennai, 
where most of the work was done. B.C. would also like to thank A. K. Kapoor,
V. Srinivasan, and S. Chaturvedi for helpful discussions.\\

\section*{\bf {A. Formulae for relevant integrals}}

\begin{enumerate}

\item

To do the three dimensional 
integrals of the nonplanar piece, the following identities in polar 
coordinates 
are useful. The angular integration can be done using,\\ 

\begin{equation}
\int _{0}^{\pi} d\phi~e^{ir|p|cos(\phi)}~sin^{n-2}(\phi) ~ =~
\frac {\Gamma(\frac{n-2}{2}) \sqrt {\pi} }{(\frac{r|p|}{2}^{\frac{n-2}{2} }) }
~J_{\frac{n-2}{2}}(r|p|) ~~~,
\end{equation}

where ~$J_{\frac{n-2}{2}}$~ is the Bessel's function of order 
~$\frac {n-2}{2}$~.  
For the temperature independent part of the nonplanar piece, 
the integration of the radial part can be done through,\\

\begin{equation}
\int _{0}^{\infty} dr~r^{\frac {n}{2}} ~(c^2 + r^2)^{m}
~J_{\frac{n-2}{2}}(r|p|) ~=~ (2/|p|)^{m+1}~\frac { c^{\frac{n}{2} +m} }
{\Gamma(-m)}~K_{\frac{n}{2} +m}(c|p|) ~~~,
\end{equation}

whereas for the temperature dependent part, we use,\\

\begin{equation}
\int _{0}^{\infty} dr~r~J_{0}(pr)~\frac { e^{-a \sqrt {r^2 + M^2} } }
{\sqrt {r^2 + M^2}} ~=~ \frac { e^{-M \sqrt {p^2 + a^2} } }
{\sqrt {p^2 + a^2}} ~~~.
\end{equation}

\item

The following properties and identities involving the modified Bessel's 
function of the second kind,  ${\mathit {K}}_{\nu}(z)$ ~ 
have been used in the text:

\begin{enumerate}

\item

\begin{equation}
\int _{0}^{\infty} d~x ~x^{\nu -1} ~e^{ \frac {\beta}{x} - \gamma ~x } = 2
( \frac {\beta}{\gamma} )^{\frac {\nu}{2} } {\mathit {K}}_{\nu}(2 \sqrt
\beta \gamma) ~~~,
\end{equation}

\item

\begin{eqnarray*}
{\mathit {K}}_{ \underline + \frac {1}{2} }(z) &=& \sqrt {\frac {\pi}{2z}
} ~ e^{-z},\\
{\mathit {K}}_{\underline +\frac {3}{2} }(z) &=& \sqrt {\frac {\pi}{2z} }
~(1 + \frac {1}{z} )~ e^{-z},\\
\end{eqnarray*}

\item

\begin{equation}
\frac {d}{dz}{\mathit {K}}_{0}(z) ~=~ - {\mathit {K}}_{1}(z) ~~~.
\end{equation}

\item

For small arguments, i.e., ~$z \rightarrow 0$~ 

\begin{equation}
{\mathit {K}}_{1}(z)~ \approx ~ \frac {1}{z} ~~~,
\end{equation}

and for large arguments, i.e., the asymptotic expansion to leading order is,

\begin{equation}
{\mathit {K}}_{1}(z)~ \approx ~\sqrt {\frac {\pi}{2z}} ~ e^{-z} ~~~.
\end{equation}

\end{enumerate}
\end{enumerate}


\end{document}